\documentclass[12pt]{article}  

\usepackage{graphicx,amsmath}
\usepackage{units}

\newlength{\dinwidth}                                                    
\newlength{\dinmargin}                                                    
\def\lapproxeq{\lower .7ex\hbox{$\;\stackrel{\textstyle                                                    
<}{\sim}\;$}}                                                    
\def\gapproxeq{\lower .7ex\hbox{$\;\stackrel{\textstyle                                                    
>}{\sim}\;$}}                                                    
\def\be{\begin{equation}}                                                    
\def\ee{\end{equation}}                                                    
\def\bea{\begin{eqnarray}}                                                    
\def\eea{\end{eqnarray}}

\begin{document}                                                    
\titlepage                                                    
\begin{flushright}                                                    
IPPP/10/40   \\
DCPT/10/80 \\                                                    
\today \\                                                    
\end{flushright}                                                    
                                                    
\vspace*{2cm}                                                    
                                                    
\begin{center}                                                    
{\Large \bf The possibility that the triple-Pomeron coupling\\}                                                    
\vspace*{0.3cm}
{\Large \bf   vanishes at $q_t$=0}                                                    
                                                    
\vspace*{1cm}                                                    
E.G.S. Luna$^{a}$, V.A. Khoze$^{b,c}$, A.D. Martin$^b$ and M.G. Ryskin$^{b,c}$ \\                                                    
                                                   
\vspace*{0.5cm}
$^a$ Instituto de F\'{\i}sica e Matem\'atica, Universidade Federal de Pelotas,
Caixa Postal 354, CEP 96010-900, Pelotas, RS, Brazil  \\                                                  
$^b$ Institute for Particle Physics Phenomenology, University of Durham, Durham, DH1 3LE \\                                                   
$^c$ Petersburg Nuclear Physics Institute, Gatchina, St.~Petersburg, 188300, Russia            
\end{center}                                                    
                                                    
\vspace*{2cm}                                                    
                                                    
\begin{abstract}                                                    

We study the case when the triple-Pomeron vertex is assumed to have a
vectorial form, that is, the amplitude of high-mass diffractive dissociation vanishes  as $V\propto \vec q_t\cdot\vec e$ as $q_t\to 0$. We find that the available data in the triple-Reggeon region may be well described in such a `weak' coupling scenario, providing that absorptive effects are taken into account.  We compare this weak (vector) coupling scenario with the strong and weak (scalar) coupling scenarios. Corresponding predictions are presented for an LHC energy of 14 TeV.

\end{abstract}    
%\newpage           

\section{Introduction}

The energy behaviour of the scattering amplitude
may be consistently described by two different scenarios for the asymptotic regime \cite{bh75}.
One is called the {\it weak coupling} of the Pomerons. In this case,
at very high energy, $\sqrt{s}$, the cross sections tend to the
universal constant value
\be
\sigma_{\rm tot} \to {\rm constant}~~~~~~{\rm as}~~~~s \to \infty.
\ee
In order not to violate unitarity, the triple-Pomeron coupling must vanish with vanishing transverse momentum, $q_t$, transferred through the Pomeron \cite{GribMYF}
\be
g_{3P}~\propto~ q^2_t~~~~~~{\rm as}~~~~q_t \to 0.
\label{eq:weak}
\ee
Another possibility is called the {\it strong coupling} scenario \cite{GribM}. Here, at a
very high energies, the cross sections grows as 
\be
\sigma_{\rm tot}\propto {({\rm ln}~s)}^\eta ~~~~~{\rm with}~~~~~ 0<\eta \leq 2,
\ee 
and the bare vertex 
\be
g_{3P}|_{q_t \to 0}~~\to~~{\rm constant}.
\label{eq:strong}
\ee

The present data are usually described within the
Froissart-like limit of the second scenario (with $\eta=2$).
However to reach asymptotics we need {\bf very} high energy --
the energy at which the slope of the elastic amplitude,
$B=B_0+\alpha'_P{\rm ln}(s)$ is dominated by the second term, that is
when $\alpha'_P{\rm ln(s)} \gg B_0$. This is far beyond the energies
available at present. Another possibility, to distinguish between the {\it
weak} and {\it strong} approaches, is to study the $q_t$ dependence of the
bare triple-Pomeron vertex \cite{AKLR}. Thus, it is important to extract the {\it
bare} vertex before its behaviour is affected by absorptive
corrections.

In Ref. \cite{LKMR} we analysed the data in the triple-Pomeron region accounting for absorptive effects in the framework of a two-channel eikonal, see also \cite{kp}.   That is, we performed a triple-Regge analysis of the available $d^2\sigma/dtd\xi$ data for $pp \to pX$ and $\bar{p}p \to \bar{p}X$ (where $\xi=M^2/s$ and $M$ is the mass of system $X$), allowing for screening (absorptive) effects. To be precise, we fitted the CERN-ISR\footnote{We chose a subset of the ISR data which is sufficient to fully describe their $t$ and $\xi$ dependence.} \cite{jcma}, FNAL fixed-target \cite{rlc} and Tevatron \cite{fa} data for $pp \to pX$ and $\bar{p}p \to \bar{p}X$. The differential distributions for the FNAL fixed target and Tevatron experiments can be found in Ref.~\cite{GM}. Both the `strong' and `weak' coupling scenarios were considered. 
We found that the data favoured the `strong' coupling scenario \cite{LKMR}. 

However, there is another possibility which should be studied. In the `weak' coupling scenario considered in \cite{LKMR} a `scalar' form of the vanishing of the coupling as $q_t \to 0$, (\ref{eq:weak}), was assumed. As was pointed out by V.N. Gribov \cite{VNG}, it is natural to have a vector form of the triple-Pomeron vertex which vanishes as $q_t \to 0$. Indeed in Feynman diagrams for the $pp \to p+X$ amplitude we never deal with $\sqrt{q^2}$, but rather with vector ${\vec q}$ multiplied by some vector ${\vec e}$ that characterizes the final state $X$.
Thus we may have a weak coupling in which the vertex of $p \to X$ dissociation has the vectorial form\footnote{Note that a vector form of the vertex for hadron production by Pomeron fusion, $V_{PP \to {\rm hadrons}}=g_{\rm had}(\vec{q}_1,\vec{q}_2)$, was considered in \cite{alrk}.}
\be
V ~\propto ~\vec{q}_t \cdot \vec{e}~~~~~({\rm that ~is}~~\sqrt{g_{3P}}\propto\vec{q}_t).
\label{eq:pol}
\ee
An example is photon-exchange. In this case the vector form of the $p \to X$ vertex comes from current conservation (gauge invariance), and is clearly seen in the Weizs\"{a}cker-Williams approach \cite{ww}, where the polarisation vector of a Coulomb-like photon, $\vec{\epsilon}$, is replaced by $\vec{\epsilon}=\vec{q}_t /x$. Here, the  (Coulomb) photon plays the role of Pomeron exchange and `$x$' is the momentum fraction transferred through the photon (or Pomeron). 

The `polarisation' structure, (\ref{eq:pol}), will change the predicted cross sections, since it leads to different screening corrections. However, can this `vectorial weak' scenario describe the data in the triple-Pomeron domain?

\section{Screening corrections in the triple-Regge formalism}
\begin{figure} 
\begin{center}
\includegraphics[height=4cm]{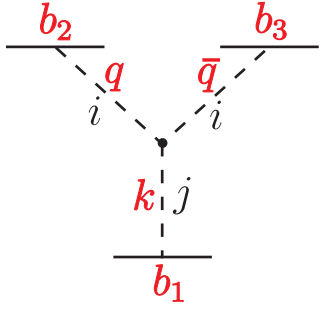}
\caption{\sf A schematic diagram showing the notation of the impact parameters arising in the calculation of the screening corrections to the $iij$ triple-Regge diagram. The conjugate momenta to $b_1,b_2,b_3$ are $k_t,q_t,\bar{q}_t$. If $k_t=0$, then $\bar{q}_t=q_t$.}
\label{fig:3Rb}
\end{center}
\end{figure}
If, for the moment, we neglect the screening correction, then the $iij$ triple-Regge diagram of Fig. \ref{fig:3Rb} gives the contribution
\be
\frac{M^2 d\sigma}{dtdM^2}~=~\beta_j(0)\beta_i^2(t)g_{iij}(t)\left(\frac{s}{M^2}\right)^{2\alpha_i(t)-2}\left(\frac{M^2}{s_0}\right)^{\alpha_j(0)-1},
\ee
where $\beta_i$ is the coupling of Reggeon $i$ to the proton, $\alpha_i(t)$ is the trajectory of Reggeon $i$, and $M$ is the mass of the system $X$ produced by proton dissociation (in Fig. \ref{fig:3Rb} this system is described by Reggeon $j$).
Let us first recall the `strong' coupling case, (\ref{eq:strong}), that we studied in \cite{LKMR}. For this we used a simple exponential parametrisation of the triple-Regge vertices
\be
g_{iij}(t)~=~g_{iij}(0)~{\rm exp}(b'_{iij}(q^2+\bar{q}^2-k^2_t)),
\ee
where the momenta are defined in Fig. \ref{fig:3Rb}, and where $q^2=t_{\rm min}-q^2_t/x_L$.

Screening effects were then included by working in impact parameter, $b$,  space and using suppression factors of the form exp($-\Omega(b)$). Recall that in the eikonal approach the elastic amplitude has the form
\be
T_{\rm el}~=~1-e^{-\Omega/2},
\ee
where the opacity
\be
\Omega=\sum_{i=P,R} \Omega_i~~~~{\rm with}~~\Omega_P=\beta_P^2 (0)\left(\frac{s}{s_0}\right)^{\alpha_i(0)-1}\frac{{\rm exp}(-b^2/4B_P)}{B_P},
\ee
where $~B_P=r_P+\alpha_P^\prime {\rm ln}(s/s_0)~$ is the $t$-slope of the Pomeron exchange amplitude. A similar expression holds for the Reggeon opacity $\Omega_R$.

To determine the $q_t$ or $t$ dependence we took the Fourier transforms with respect to the impact parameters specified in Fig. \ref{fig:3Rb}. We then obtained\footnote{Note that $e^{i\vec{k}_t \cdot \vec{b}_1}=1$ as $k_t=0$.}
\be
\frac{M^2 d\sigma}{dtdM^2}~=~A\int\frac{d^2b_2}{2\pi}e^{i\vec{q}_t \cdot \vec{b}_2} F_i(b_2)\int\frac{d^2b_3}{2\pi}e^{i\vec{q}_t \cdot \vec{b}_3} F_i(b_3)\int\frac{d^2b_1}{2\pi} F_j(b_1),
\label{eq:3Rb}
\ee
where
\be
F_i(b_2)~=~\frac{1}{2\pi \beta_i(q_t=0)}\int d^2q_t \beta_i(q_t)\left(\frac{s}{M^2}\right)^{-\alpha^\prime_i q^2_t} e^{b^\prime_{iij}q^2}e^{i\vec{q}_t \cdot \vec{b}_2},
\label{eq:Fi}
\ee
\be
F_j(b_1)~=~\frac{1}{2\pi \beta_j(k_t=0)}\int d^2k_t \beta_j(k_t)\left(\frac{M^2}{s_0}\right)^{-\alpha^\prime_j k^2_t} e^{-b^\prime_{iij}k^2_t},
\label{eq:Fj}
\ee
and where the $q_t$-independent factors are collected in $A$
\be
A~=~\beta_j(0)\beta_i^2(0)g_{iij}(0)\left(\frac{s}{M^2}\right)^{2\alpha_i(t_{\rm min})-2}\left(\frac{M^2}{s_0}\right)^{\alpha_j(0)-1}.
\label{eq:A}
\ee
These equations are relevant for the strong triple-Pomeron scenario, see (\ref{eq:strong}). In Ref. \cite{LKMR} we also studied the `scalar' weak triple-Pomeron coupling ansatz, (\ref{eq:weak}).  To do this we included a factor $q_t$ in the integrand of the expression (\ref{eq:Fi}) for $F_i(b_2)$ when $i,j=P$, and also $\bar{q}_t~(=q_t)$ in the analogous formula for $F_i(b_3)$.

To obtain the screening corrections, for a single-channel eikonal, we then included in the integrands on the right-hand side of (\ref{eq:3Rb}) the factors
\be
{\rm exp}(-\Omega(\vec{b}_2-\vec{b}_1)/2)~{\rm exp}(-\Omega(\vec{b}_3-\vec{b}_1)/2)~\equiv~S(\vec{b}_2-\vec{b}_1)~S(\vec{b}_3-\vec{b}_1).
\ee
That is, we computed
\be
\left. \frac{M^2 d\sigma}{dtdM^2}\right|_{iij}~=~A\int\frac{d^2b_1}{2\pi} F_j(b_1) |I(b_1)|^2,
\label{eq:3Rbscreen}
\ee
where $I$ is given by
\be
I(b_1)~\equiv~\int\frac{d^2b_2}{2\pi}e^{i\vec{q}_t \cdot \vec{b}_2} F_i(b_2) S_i(\vec{b}_2-\vec{b}_1).
\label{eq:Sj}
\ee
Here $q_t$ is the transverse momentum of the outgoing proton, which is now the sum of the transverse momentum of the Pomeron coupling to the triple-Pomeron vertex and that of the screening Pomeron.

These were the `strong' and `weak' scenarios studied in detail in \cite{LKMR}. As mentioned above, we found that the data favoured the `strong' coupling scenario. Our concern here is the possibility that we may have a `weak' triple-Pomeron coupling with a vectorial structure. That is the coupling in Fig. \ref{fig:3Rb} is of the form
\be
g_{3P}(t)~=~g_{3P}(0)~{\rm exp}(b'_{iij}(q^2+\bar{q}^2-k^2_t))\cdot \delta_{\mu,\nu}q_{t\mu}\bar{q}_{t\nu}.
\ee
The final product means that an additional factor $\vec{q}_t$ will occur in the integrand of (\ref{eq:Fi}). 
The presence of a vectorial $\vec{q}_t$ means that after the angular integration in (\ref{eq:Fi}), we now obtain the Bessel function $J_1(q_tb)$, and not $J_0(q_tb)$ as was found in \cite{LKMR}. Also, in the impact parameter, $b$, representation the amplitude corresponding to (\ref{eq:Fi}) takes a vector form
\be
\vec{F}_i(\vec{b})~=~\vec{b} f(b).
\ee  
Due to the factor $J_1(q_tb)$, the amplitude vanishes at $b=0$, where the screening effect, exp$(-\Omega)$, is at its maximum. Thus, we now anticipate a weaker screening for the triple-Pomeron term. Moreover, as the amplitude now has vector form (analogous to that occurring in the $\pi\pi P$ contribution discussed in Section 4.3 of \cite{LKMR}) we have to consider the components $I_x$ and $I_y$ separately. That is $I$ of (\ref{eq:Sj}) is given by
\be
|I|^2=|I_x|^2+|I_y|^2,
\label{eq:Ixy}
\ee
where it is convenient to direct $x$ along $\vec{q}_t$.

The generalisation of the formalism to a two-channel eikonal is straightforward. It has been presented in Section 4.2 of \cite{LKMR}.

\section{Results}
\begin{figure} 
\begin{center}
\includegraphics[height=15cm]{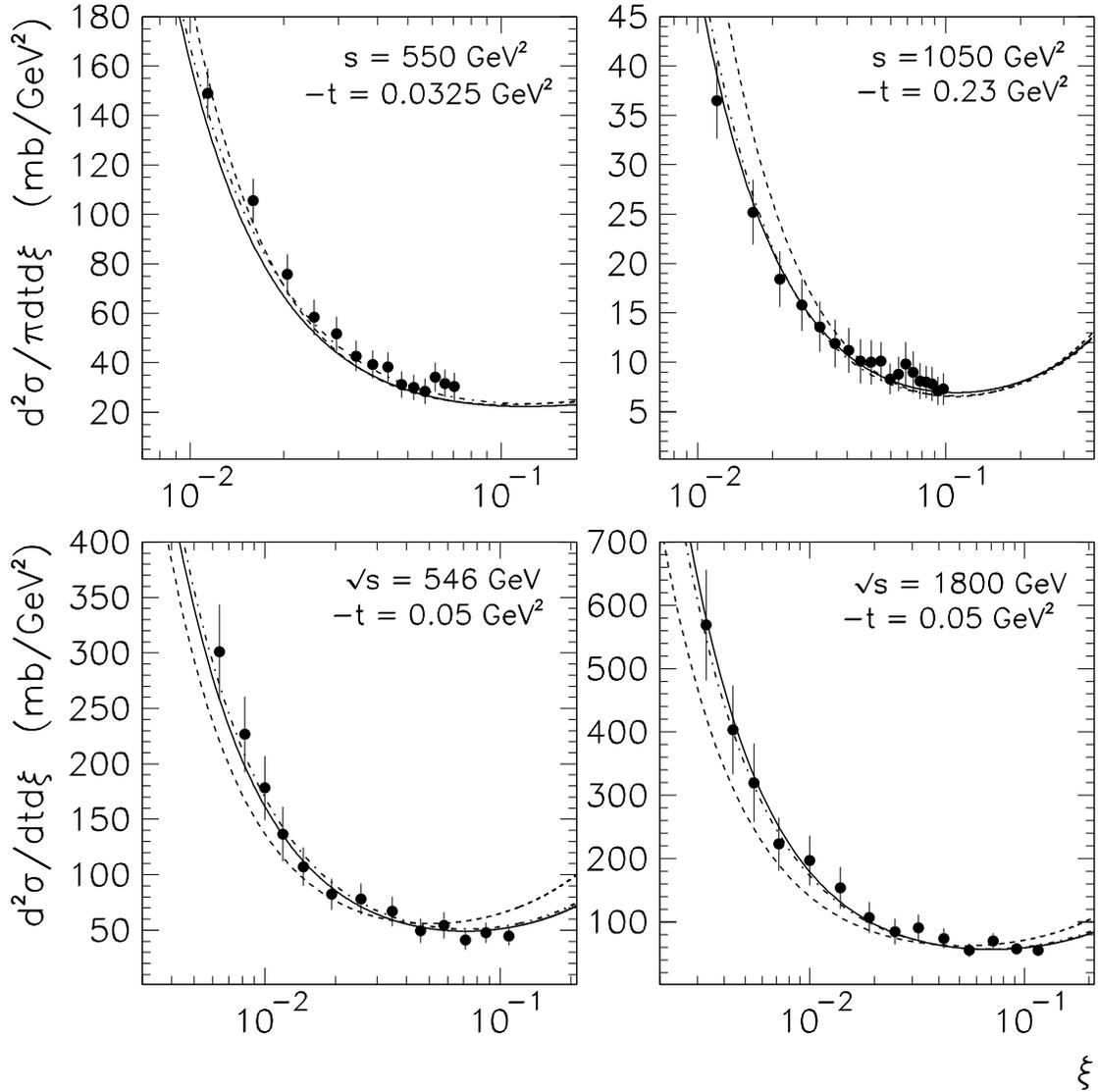}
\caption{\sf The description of a sample of the $d^2\sigma/dtd\xi$ cross section data that are fitted  using the weak-vector (continuous curves), weak-scalar (dashed curves) and strong (dot-dashed) triple-Pomeron coupling ansatzes. ($\xi\simeq M^2/s$). Here, the curves corresponding to the (strong, weak) coupling fits of the FNAL data have been normalised (down, up) by 15$\%$ at $\sqrt{s}=$ 546 GeV and by 10$\%$ at $\sqrt{s}=$ 1800 GeV, to allow for the normalisations found for these data in the respective fits.}
\label{fig:weak}
\end{center}
\end{figure}
\begin{figure} 
\begin{center}
\includegraphics[height=17cm]{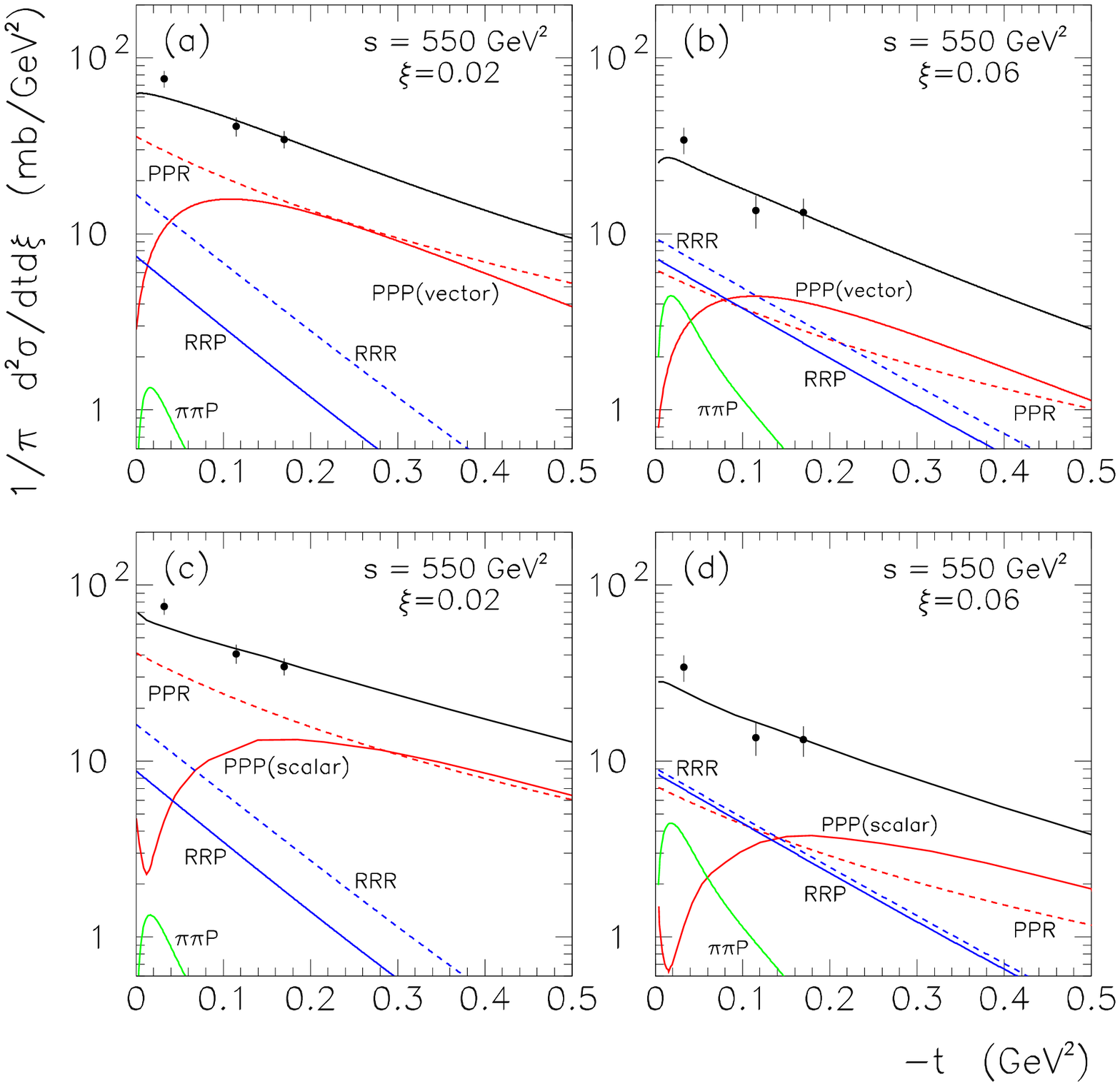}
\caption{\sf  Comparison of the $t$-dependence of $d^2\sigma/dtd\xi$ at $\xi=0.02,~0.06$ and $s=550 ~{\rm GeV}^2$ obtained using the weak-vector triple-Pomeron coupling, with that corresponding to the weak-scalar triple-Pomeron coupling used in \cite{LKMR}, together with the data available at these kinematic values.}
\label{fig:tdep2}
\end{center}
\end{figure}
\begin{figure} 
\begin{center}
\includegraphics[height=17cm]{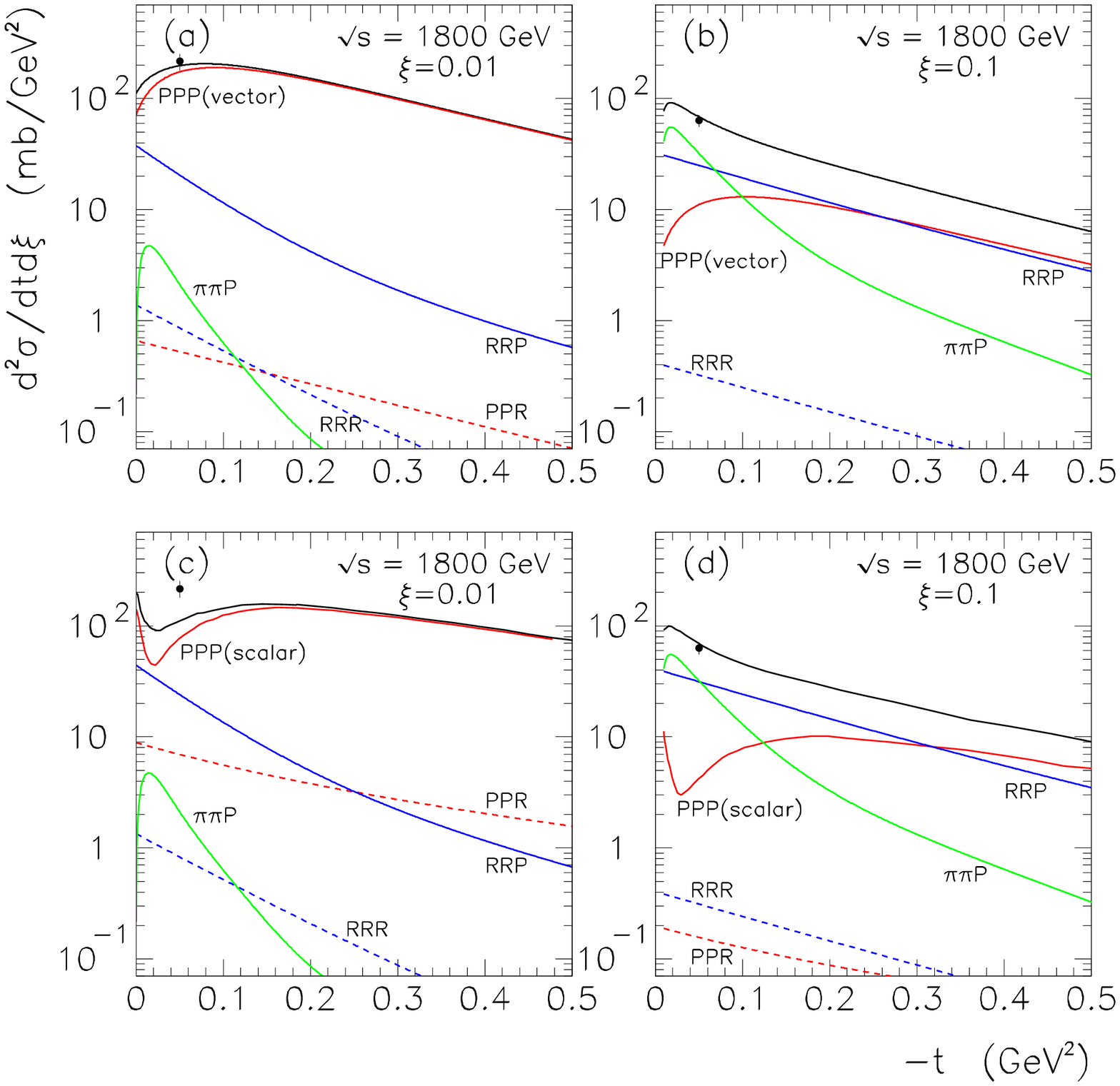}
\caption{\sf  Comparison of the $t$-dependence of $d^2\sigma/dtd\xi$ at $\xi=0.01,~0.1$ and $\sqrt{s}=1800 ~{\rm GeV}$ obtained using the weak-vector triple-Pomeron coupling, with that corresponding to the weak-scalar triple-Pomeron coupling used in \cite{LKMR}, together with the data available at these kinematic values.}
\label{fig:tdep21800}
\end{center}
\end{figure}

\begin{figure} 
\begin{center}
\includegraphics[height=14cm]{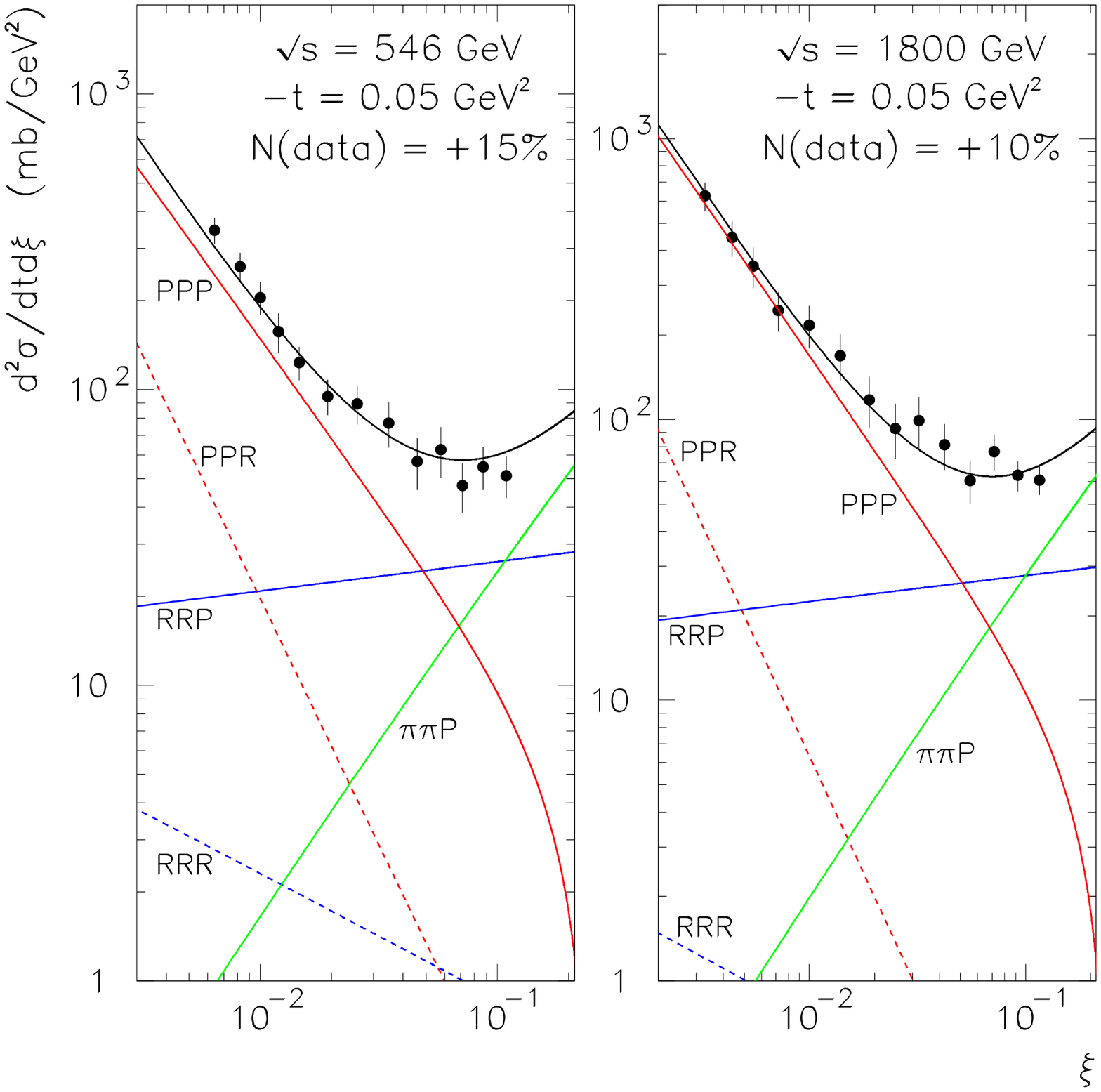}
\caption{\sf The description of the $d^2\sigma/dtd\xi$, measured in the collider experiments at FNAL \cite{rlc,fa,GM}, obtained in the weak-vector  triple-Pomeron coupling fit. The individual triple-Regge contributions are also shown.}
\label{fig:Tev}
\end{center}
\end{figure}

As can be seen from Figs. \ref{fig:weak}$-$\ref{fig:Tev}, the presently available data in 
the triple-Regge region are well described assuming that the triple-Pomeron vertex has a vectorial `weak' form. Fig. \ref{fig:weak} also shows the results of two previous fits, taken from Ref. \cite{LKMR}, which were obtained with the assumption that the triple-Pomeron vertex had, first, a `strong' coupling and, then, a scalar `weak' coupling. All three fits used the same data sets. Recall that these three scenarios differ in how the triple-Pomeron coupling vanishes as $q_t \to 0$:
$$ {\rm weak~vector~coupling}:~~~\sqrt{g_{3P}} \propto \vec{q_t} $$
$$ {\rm weak~scalar~coupling}:~~~\sqrt{g_{3P}} \propto q_t $$
$$ ~~~~~~ ~~~{\rm strong~coupling}:~~~~~~~~~\sqrt{g_{3P}} \propto {\rm constant}. $$
Table 1 compares the values of parameters obtained for the vectorial weak scenario with those obtained in Ref. \cite{LKMR} for the strong and scalar weak scenarios.  Note that for the case of vectorial `weak' coupling we get  
 practically the same $\chi^2$ as in the 
more popular `strong' coupling scenario. The secondary Reggeon contributions, that is the $RRP$, $RRR$ and $PPR$ terms, coincide,
within the error bars, in all three scenarios.
\begin{table}[htb]
\begin{center}
\begin{tabular}{|c|c|c|c|}\hline
 & strong & weak-scalar & weak-vector \\ \hline
$g^S_{3P}$  &   $0.44 ~\pm~0.05 $ &-& -  \\  
$g^W_{3P}$  &   - &$3.0 ~\pm~1.2 $& $3.1 \pm 0.5$ \\ 
$b^{\prime W}_{PPP}$  &   - &$1.15 ~\pm~0.3 $&  $0.9\pm 0.4$ \\ 
$g_{PPR}$  &   $0.75 ~\pm~0.10 $ &   $0.76 ~\pm~0.15 $&   $0.66 ~\pm~0.11 $  \\
$b^{\prime W}_{PPR}$  &   - &$1.4 ~\pm~1.7 $& $0.8\pm 0.8$  \\ 
$g_{RRP}$  &   $1.1 ~\pm~0.3  $ &   $1.3 ~\pm~0.5  $&   $1.1 ~\pm~0.4  $ \\
$g_{RRR}$  &   $2.6 ~\pm~1.0  $ &   $2.9 ~\pm~1.4  $&   $3.0 ~\pm~1.1  $  \\ \hline
$\chi^2$/DoF & 0.83 & 1.40 & 0.86 \\ \hline
\end{tabular}
\end{center}
\caption{\sf The values of the {\it ``bare''} triple-Regge couplings $g_{iij}(0)$ of (\ref{eq:A}), and slopes $b'_{iij}$ of (\ref{eq:Fi}, \ref{eq:Fj}), obtained in the three optimum fits to the $d^2\sigma/dtd\xi$ data.  GeV units are used; so, for example, the couplings $g_{3P}$ have units of ${\rm GeV}^{-1}$.
The parameters for the strong and weak-scalar fits are taken from Ref. \cite{LKMR}. Recall that all the slopes $b'_{iij}$ are set to zero, except for those of the $PPP$ and $PPR$ vertices in the weak coupling fit.}
\end{table}

\begin{figure} [t]
\begin{center}
\includegraphics[height=12cm]{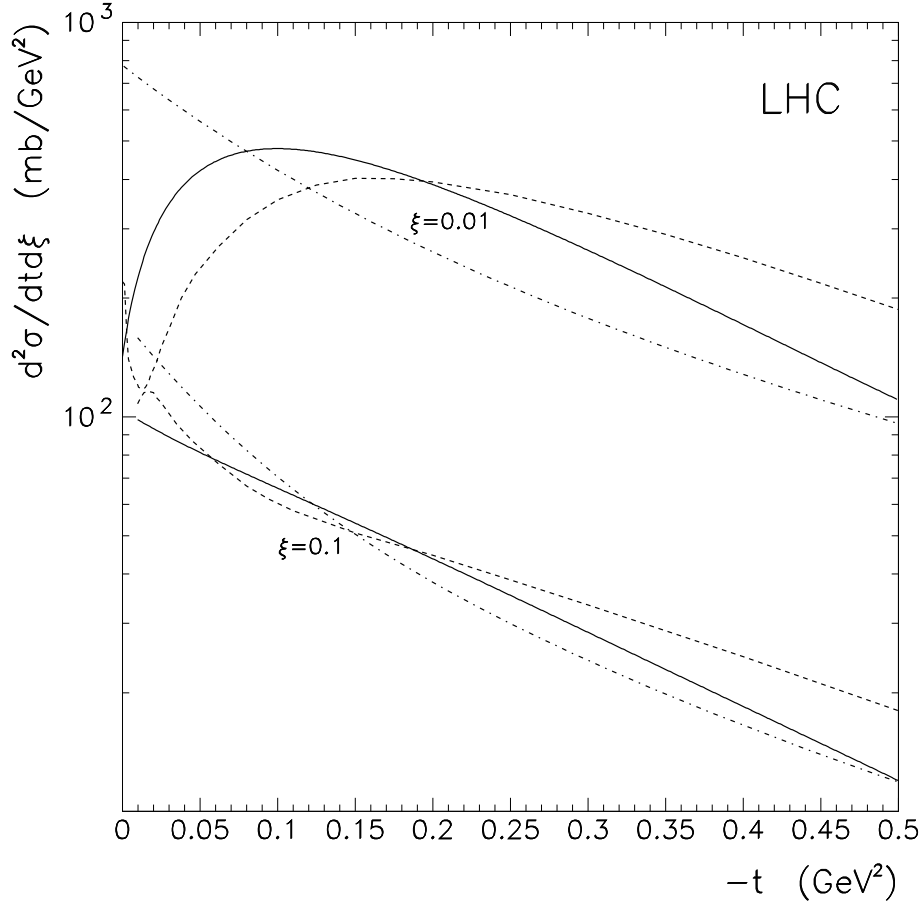}
\caption{\sf The predictions for the $t$-dependence of the $d^2\sigma/dtd\xi$ at $\xi=0.01,~0.1$ and $\sqrt{s}=14$ TeV obtained using the weak-vector (continuous curves), weak-scalar (dashed) and strong (dot-dashed) triple-Pomeron coupling scenarios.}
\label{fig:tdepLHC}
\end{center}
\end{figure}

Unlike that for the weak-scalar, the weak-vector triple-Pomeron contribution has no dip at very low $|t|\sim 0.02$ GeV$^2$ (see Figs. \ref{fig:tdep2}, \ref{fig:tdep21800} and \ref{fig:tdepLHC}). The  {\it vector} contribution vanishes  at $q_t=0$ since we have no direction.

% This new form of the $PPP$ curve may be explained as follows.
%As $t\to 0$ (that is, $q_t\to 0$) the `vector' contribution vanishes since we have no direction. 
%Next, instead of a narrow dip we have a relatively shallow minimum at $|t|\sim 0.12\ -\  0.2$ GeV$^2$ (at the LHC $-$ CERN-ISR energies) caused by the absorptive corrections to the original triple-Pomeron amplitude with the dipole-like form of the nucleon-Pomeron vertex $\beta_P(t)$
% (see eq.(18) of Ref.\cite{LKMR}). 
%the two vector components $I_x$ and $I_y$ of (\ref{eq:Ixy}) have dips
%at different values of $t$. Therefore the sum $|I|^2$ does not have
%a deep dip, but instead relatively shallow minima and maxima. 
%The 
More rapid decrease of the $PPP$ contribution at large $\xi>0.1$,
 seen in Fig. \ref{fig:Tev}, is due to the longitudinal part of 
the momentum transfer 
\be
-t=q^2_t+q^2_\|=(q^2_t+\xi^2 m^2_p)/(1-\xi).
\ee
 In the vectorial `weak' scenario the triple-Pomeron contribution vanishes as $q_t\to 0$, while for fixed $t$, the value of $q_t$ decreases as $\xi$ increases. 
Measurements of $d^2\sigma/dtd\xi$ at the LHC,
especially at small $\xi\sim 0.01$, should be able to distinguish between the three scenarios, see Fig. \ref{fig:tdepLHC}.

\section{Discussion}

Recall that in perturbative QCD, the leading-order BFKL triple-Pomeron vertex does not vanish, but takes a non-zero constant value at $q_t=0$ . This result corresponds to interactions at very small distances. However, at larger distances, relevant for $q_t \to 0$, the absorptive effects caused by enhanced multi-Pomeron diagrams could modify the perturbative QCD result leading to a vanishing triple-Pomeron coupling as $q_t \to 0$. This vanishing behaviour also looks natural if we bear in mind confinement, which does not allow colour-induced interactions at large distances.
From this point of view it looks encouraging that the present triple-Regge data may be well described within the vectorial
`weak' approach, after accounting for absorptive corrections. 

The only minor problem is the inelastic $J/\psi$ diffractive 
photoproduction observed at HERA \cite{jpsi}. Due to the small cross section of the $J/\psi$-proton interaction, the absorptive effects in this case are very small.
 The ratio, $r$, of the cross section with proton dissociation, $\gamma p\to J/\psi + Y$, integrated over the mass region $M_Y<30$ GeV, to that of  `elastic' photoproduction,
$\gamma p\to J/\psi + p$, was measured as a function of momentum transfer $t$. At the smallest value of $-t=0.2$ GeV$^2$ measured by ZEUS\cite{jpsi}, this ratio $r=0.4\pm 0.1$ \cite{jpsi,jel}. On the other hand, using the parameters of our `weak-vector' fit from Table 1, we
predict a smaller value 
 $$r=r_{PPP}+r_{PPR}\simeq 0.16+0.08\ =0.24.$$ 
Here $r_{PPP}$ 
and $r_{PPR}$ denote the contributions of the $PPP$ and $PPR$ terms respectively.
This should be compared to the predictions $r \simeq 0.16+0.12=0.28$ and $r \simeq 0.14+0.07=0.21$ obtained using the strong and `weak-scalar' fits of Ref. \cite{LKMR}. Indeed, in \cite{LKMR} we concluded that the weak-scalar triple-Pomeron coupling was disfavoured both by the fit to the triple-Regge data and by the $J/\psi$ data.  On the other hand, the triple-Regge data can be reasonably accommodated by the weak-vector coupling, although this triple-Pomeron coupling is still a bit disfavoured, in comparison with the strong coupling regime, by the $J/\psi$ data.
 Unfortunately, (a) we have no data at smaller $|t|$ and (b) the cross section of inelastic $\gamma p\to J/\psi+Y$ diffractive photoproduction
has 
% was (for some reason) 
never been published in a 
%scientific
 journal. Thus, these data cannot be considered as a strong argument against the `weak-vector' coupling scenario.

It would be very interesting to measure the $t$ dependence of high-mass diffractive dissociation, $d^2\sigma/dtd\xi$, at the LHC in order to choose between the possible vectorial `weak' and `strong'
coupling asymptotic regimes.

\section*{Acknowledgements}

MGR and EGSL thank the IPPP at the University of Durham for hospitality.
The work was supported  by the Russian State grant RSGSS-3628.2008.2.

\end{document}